\documentclass[a4paper,journal,draftcls,10pt,onecolumn]{IEEEtran}
\usepackage{cite}

\ifCLASSINFOpdf
\else
\fi

\usepackage{multirow}
\usepackage{graphicx}
\usepackage{mathrsfs}

\newtheorem{theorem}{\bf{Theorem}}

\begin{document}
%
\title{Impact of Preference and Equivocators on Opinion Dynamics with Evolutionary Game Framework}
%
%
%

\author{{Xinyang~Deng}{\#},
        {Zhen~Wang}{\#},
        {Qi~Liu},
        {Yong~Deng}{*},
        and~{Yu~Shyr}
\thanks{X. Deng, Z. Wang and Y. Deng are with the School of Computer and Information Science, Southwest University, Chongqing, 400715, China.}
\thanks{X. Deng, Q. Liu and Y. Shyr are with the Center for Quantitative Sciences, Vanderbilt University School of Medicine, Nashville, TN, 37232, USA.}
\thanks{Z. Wang is with the Interdisciplinary Graduate School of Engineering Sciences, Kyushu University, Kasuga-koen, Kasuga-shi, Fukuoka 816-8580, Japan.}
\thanks{Q. Liu is with the Department of Biomedical Informatics, Vanderbilt University School of Medicine, Nashville, TN, 37232, USA.}
\thanks{Y. Deng is with the School of Engineering, Vanderbilt University, Nashville, TN, 37235, USA.}
\thanks{\# These authors contribute equally.}
\thanks{* Corresponding author: Y. Deng, School of Computer and Information Science, Southwest University, Chongqing, 400715, China. E-mail: prof.deng@hotmail.com; ydeng@swu.edu.cn.}}

%
%

\markboth{Journal of \LaTeX\ Class Files,~Vol.~13, No.~9, September~2014}%
{Shell \MakeLowercase{\textit{et al.}}: Bare Demo of IEEEtran.cls for Journals}
%



\maketitle

\begin{abstract}
Opinion dynamics, aiming to understand the evolution of collective behavior through various interaction mechanisms of opinions, represents one of the most challenges in natural and social science. To elucidate this issue clearly, binary opinion model becomes a useful framework, where agents can take an independent opinion. Inspired by the realistic observations, here we propose two basic interaction mechanisms of binary opinion model: one is the so-called BSO model in which players benefit from holding the same opinion; the other is called BDO model in which players benefit from taking different opinions. In terms of these two basic models, the synthetical effect of opinion preference and equivocators on the evolution of binary opinion is studied under the framework of evolutionary game theory (EGT), where the replicator equation (RE) is employed to mimick the evolution of opinions. By means of numerous simulations, we show the theoretical equilibrium states of binary opinion dynamics, and mathematically analyze the stability of each equilibrium state as well.
\end{abstract}


%
\IEEEpeerreviewmaketitle

\section{Introduction}
In realistic life, there are large amount of opinion interactions on many issues of interest among social individuals. To understand the evolution and formation of opinions, opinion dynamics has provided a useful framework in theoretical and experimental research areas  \cite{lorenz2007continuousIJMPC,acemoglu2011opinion}. Generally speaking, the opinion can be divided in two types:  continuous opinion and discrete opinion, both of which have been extensively investigated and further extended to more scenarios, such as DeGroot model \cite{degroot1974reaching}, discrete CODA model \cite{martins2008continuous}, Hegselmann-Krause model \cite{hegselmann2002opinion}, generalized Glauber models \cite{castellano2011irrelevance}, and so on \cite{galam2008sociophysics,deffuant2000mixing,liu2013opinionsr,di2007opinion,quattrociocchi2014opinion,Zheng2015162AMC,carletti2008birth}. While among the existing achievements, binary opinion, as one typical case of discrete fashion, has attracted particular attention due to its simplicity like \emph{yes/no}, \emph{agree/disaggre}, \emph{accept/reject}. In this sense, binary opinion model becomes a straightforward metaphor to help us understand the evolution of opinions \cite{galam2005local}.

During the past decades, the research of binary opinion has gained growing interest, both analytically and numerically. For example, from the viewpoint of statistical physics,  Sznajd model \cite{sznajd2000opinion} could be regarded as a preferential version to inspect the opinion formation by borrowing the framework of Ising model. Baker and Hague \cite{baker2008rise} extended the Sznajd model to continuous and multi-state discrete opinions. In \cite{biswas2009model}, the authors used the size of neighboring domains to quantify the social pressure, and further proposed a one-dimensional model of binary opinion. Guo et al. \cite{guo2013much,guo2012bifurcation} investigated the evolution of binary opinion on networks, where  the heterogeneity of opinion interaction and randomness of human decision were considered \cite{guo2013heterogeneity}. In \cite{chi2011binary}, the influence of noise was incorporated into the binary opinion dynamics. Biswas et al. \cite{biswas2013opinion} proposed a weighted influence model (WI model). As a standard framework to study how cooperation emerges, evolutionary game theory (EGT) \cite{smith1973the,hofbauer2003evolutionary,weibull1997evolutionary} has also been utilized to explore the evolution and formation of opinion in some studies recently \cite{cao2008mixed,ding2009coIJMPC,gargiulo2012influence}.

Besides, the evolutionary mechanism is crucial in the evolution of opinion. In EGT, replicator equation (RE) \cite{ding2010evolutionary,taylor1978evolutionary,hofbauer1998evolutionary} provides a rule to simulate the evolution of strategies in populations. Mathematically, it is equivalent to the Lotka - Volterra equations of ecology \cite{hofbauer1998evolutionary,mobilia2007phase}, which describes the dynamics of species in an interacting biological system. The RE is very appropriate to act as the role of evolutionary mechanism in opinion dynamics. On the one hand, compared with the classical game theory, the RE does not rely on any assumption of rationality, which is more close to the real situation. On the other hand, opinions have high mobility and are very easy to diffuse, which is approximatively meet the well-mixed requirement of RE. Therefore, in this paper the RE has been utilized to simulate the evolution of opinions. In the evolutionary process of opinions, each opinion is naturally regarded as a species, the spreading of opinions is analogous to the propagation of species.

Apart from the evolutionary mechanism of opinions, the interaction mechanism of opinions has also played a very important role in opinion dynamics. For example, in \cite{cao2008mixed} Cao and Li employed the battle-of-the-sexes game to model the opinion formation on networks. In \cite{ding2010evolutionary}, Ding et al. used cooperative game and minority game to describe two types of opinion interactions. Essentially, these game models belong to a family of so-called coordination game \cite{van1990tacit}. In coordination game, there are two basic interaction mechanisms. The first one is that agents can get profits by taking the same action, the second one is that agents are rewarded by taking different actions. In this paper, based on the idea of coordination game, we use two basic game models to represent the interaction between opinions. One is called BSO model in which players benefit from holding the same opinion, the other is called BDO model in which players benefit by holding different opinions. In the intuitive sense, the BSO model pays close attention on the consensus of opinions, while the BDO model encourages the diversity of opinions. In the discrete opinion dynamics, two important factors are usually considered. One is the opinion preference, the other is the existence of equivocators or centrists. For example, Ding et al. \cite{ding2010evolutionary} considered the opinion preference in cooperative and minority games, in \cite{vazquez2004ultimate,mobilia2011fixation} the authors have paid much attention on centrists in the vote model. But the synthetical effect of these two factors gets less attention especially by using the framework of EGT and RE. In this paper, in terms of these two basic models, the impact of opinion preference and equivocators on the evolution of binary opinion is studied respectively and synthetically in the framework of EGT and RE. The equilibrium states of opinion evolution have been found, and the stability of each equilibrium state has also been analysed mathematically.

In the rest of this paper, the EGT and RE is introduced first. Then, we give two basic models for interaction of opinions, and present the opinion dynamics with equivocators and preference. Finally, we will summarize our conclusions.

\section{Basic of evolutionary game theory and replicator equation}\label{SectBasicOFEGT}
Evolutionary game theory (EGT) \cite{smith1973the,smith1982evolution} was initially found  by John Maynard Smith to study the interaction among different players or populations located on various networks \cite{eguiluz2009critical,wang2012evolution,iwasa2013graduated,masuda2007participation,wang2012cooperation,masuda2012evolution,eguiluz2005cooperation,iwagami2010upstream,ohtsuki2006leading,xia2014beliefLIUXIA}. In recent years, EGT has become a paradigmatic framework to understand the emergence and evolution of cooperation among unrelated individuals \cite{nowak1992evolutionary,szabo2007evolutionary,mobilia2013evolutionary,wang2013interdependent,mobilia2010oscillatory,zimmermann2005cooperation,axelrod2006evolution,wang2013insight,burton2015payoff,santos2005scale,boccaletti2014structure,nowak2006evolutionary,deng2014impact,Deng201481JTB,shigaki2012referring}. The main idea of the EGT is to track the change of strategies' frequency of population during the evolutionary process. In EGT, the replicator equation (RE) \cite{taylor1978evolutionary,hofbauer1998evolutionary} plays a key role to determine the evolutionary process of population, which has provided a frequency-dependent evolutionary dynamics to a well-mixed population.

Assume there exists $n$ strategies in a well-mixed population. A game payoff matrix $A = [a_{ij}]$ determines the payoff of a player with strategy $i$ if he meets another player who carries out strategy $j$. The fitness of strategy $i$ is defined by:
\begin{equation}\label{Eq_fitnessfi}
f_i  = \sum\limits_{j = 1}^n {x_j a_{ij} } ,\quad i = 1, \cdots ,n,
\end{equation}
where $x_j$ is the relative frequency of strategy $j$ in the population. The average fitness of all strategies is denoted as $\phi$, which is defined by:
\begin{equation}
\phi  = \sum\limits_{i = 1}^n {x_i f_i }.
\end{equation}

The relative frequency of strategy $i$, namely $x_i$, is changed with time by this following differential equation:
\begin{equation}\label{Eq_RE}
\frac{{dx_i }}{{dt}} = x_i (f_i  - \phi ) ,\quad i = 1, \cdots ,n.
\end{equation}

Eq.(\ref{Eq_RE}) is the so-called replicator equation, which implies that the change of $x_i$ depends on the fitness of strategy $i$ and $x_i$. By solving $\frac{{dx_i }}{{dt}} = 0$, $i=1,\cdots,n$, the fixed points of this evolutionary system, denoted as $(x_1^*, \cdots, x_n^*)$, can be found. Regarding the stability of the fixed point $(x_1^*, \cdots, x_n^*)$, a theorem is usually used to verify whether the fixed point is stable or not, which is given as below.
\begin{theorem}\cite{hofbauer1998evolutionary}\label{theorem1}
Given a set of replicator equations $\frac{{dx_i }}{{dt}} = x_i (f_i  - \phi ) ,\quad i = 1, \cdots ,n$, the fixed point $p^* = (x_1^*, \cdots, x_n^*)$ is stable if all eigenvalues associated with $p^*$ are negative numbers or have negative real parts.
\end{theorem}

For more details on Theorem \ref{theorem1}, please refer to literature \cite{hofbauer1998evolutionary}. If the set of eigenvalues associated with $p^*$ consists of negative numbers and zero, reference \cite{sandholm2010population} provides a solution to judge the stability of such fixed points.

\section{Coordination game of opinions}\label{SectBasicOFBSOBDO}
In two players' opinion interaction, two cases may happen. The one is that these two players own the same opinion, the other is that the players hold different opinions. Regarding these two cases, a game model, called coordination game, can be used appropriately to describe this interaction between two players. In the classical coordination game, people will coordinate by taking the same action or taking different actions. In the coordination game of opinions, it could contain two basic models. This first one is that players benefit from holding the same opinion, which is called as BSO for simplicity in this paper. The second one is that players benefit by holding different opinions, which is called as BDO. These two basic models are abundant in the real world.

\begin{figure}[!t]
\centering
\includegraphics[width=5in]{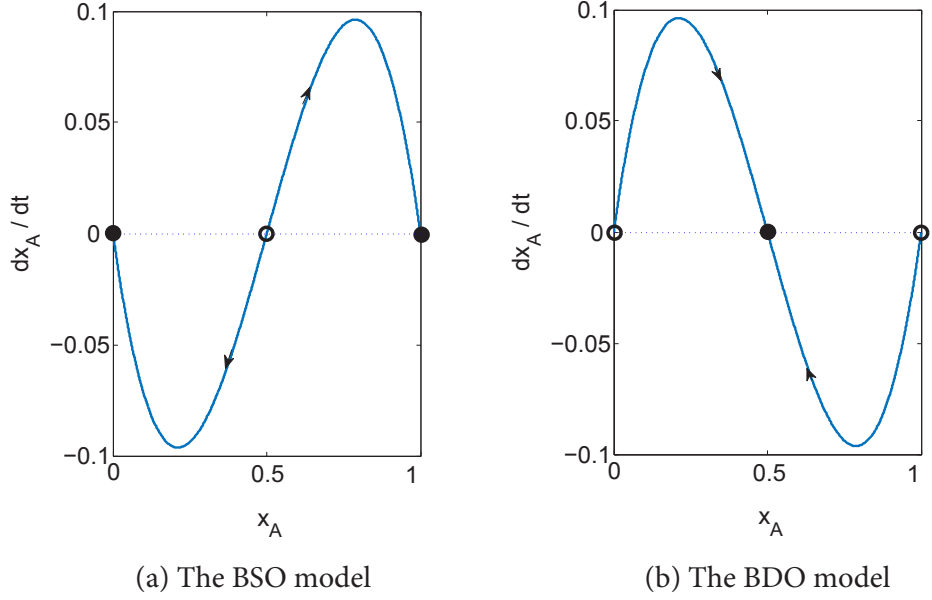}
\caption{Phase diagrams of the binary opinion's coordination games. Black circles represent stable fixed points, and white circles correspond to unstable fixed points. The arrows show the evolutionary direction.}
\label{FigPPGsCGs}
\end{figure}

The BSO model is a typical pure and symmetric coordination game, such as driving coordination game in which two drivers choose the same direction to avoid collision. The payoffs in the BSO model is shown in Eq.\ref{Eq_PT_BSO},
\begin{equation}\label{Eq_PT_BSO}
\begin{array}{*{20}c}
   {} & {\begin{array}{*{20}c}
   A & B  \\
\end{array}}  \\
   {\begin{array}{*{20}c}
   A  \\
   B  \\
\end{array}} & {\left[ {\begin{array}{*{20}c}
   1 & 0  \\
   0 & 1  \\
\end{array}} \right]}  \\
\end{array}.
\end{equation}
In the BSO model, players with the same opinion will be rewarded. As shown in Eq.\ref{Eq_PT_BSO}, if two players having the same opinion (either $A$ or $B$) meet, each of them gets the payoff of one. If two players who hold different opinions meet, each gets a payoff of zero. In this paper, ``strategy" is represented by ``opinion", ``the evolution of strategy" is represented by ``the evolution of opinion". Based on the EGT and RE, we can analyze the evolutionary process of these two opinions in the BSO model. Let the relative frequency of opinions $A$ and $B$ be indicted by $x_A$ and $x_B$, respectively, where $x_A + x_B = 1$. So the REs read
\begin{equation}\label{Eq_dxAxB}
\left\{ \begin{array}{l}
 \frac{{dx_A }}{{dt}} = x_A (f_A  - \phi ), \\
 \frac{{dx_B }}{{dt}} = x_B (f_B  - \phi ), \\
 \end{array} \right.
\end{equation}
where $f_A  = x_A  \times 1 + x_B  \times 0$, $f_B  = x_A  \times 0 + x_B  \times 1$, and $\phi  = x_A f_A  + x_B f_B$. Further, Eq.(\ref{Eq_dxAxB}) can be wrote as
\begin{equation}
\frac{{dx_A }}{{dt}} = x_A (1 - x_A )(f_A  - f_B ),
\end{equation}
namely,
\begin{equation}\label{Eq_dxA}
\frac{{dx_A }}{{dt}} = x_A (1 - x_A )(2 x_A  - 1 ).
\end{equation}
By solving $\frac{{dx_A }}{{dt}} = 0$, the fixed points $(x_A^*, x_B^*)$ of this evolutionary dynamics can be obtained easily. There exists three fixed points, $(0, 1)$, $(1, 0)$ and $(0.5, 0.5)$. According to Theorem \ref{theorem1}, the stability of each fixed point is easily known. It can be found that, fixed points $(0, 1)$, $(1, 0)$ are stable: any perturbation deviating the population from these points will induce the dynamics that restores to these fixed points. $(0.5, 0.5)$ is an unstable fixed point where any deviation from that point will move away from it as time increases. In order to have a better understanding to the stability of these fixed points, the phase diagram of Eq.(\ref{Eq_dxA}) is given in Figure \ref{FigPPGsCGs}(a). Each black circle is a stable fixed point, and each white circle is a unstable fixed point. The arrows show the evolutionary direction. As can be seen from Figure \ref{FigPPGsCGs}(a), in the BSO model these two opinions $A$ and $B$ can not be coexisting determinately. The final state of the population is determined by its initial state: the final equilibrium state is opinion $A$ if the initial frequency of opinion $A$ is bigger than that of opinion $B$; otherwise, it turns to the inverse side.

Now let us consider the opposite BDO model where individuals with different opinions can benefit. In the BDO model, the players holding different opinions are rewarded, which essentially means the diversity of opinions is encouraged. Eq.(\ref{Eq_PT_BDO}) shows the payoffs in the BDO model:
\begin{equation}\label{Eq_PT_BDO}
\begin{array}{*{20}c}
   {} & {\begin{array}{*{20}c}
   A & B  \\
\end{array}}  \\
   {\begin{array}{*{20}c}
   A  \\
   B  \\
\end{array}} & {\left[ {\begin{array}{*{20}c}
   0 & 1  \\
   1 & 0  \\
\end{array}} \right]}  \\
\end{array}.
\end{equation}
If two players with the same opinion meet, each gets a payoff of zero; If they hold different opinions, each gets a payoff of one. Similarly, in terms of the replicator equation, the relative frequency of opinion $A$, $x_A$, is changed as time increases
\begin{equation}\label{Eq_xABDO}
\frac{{dx_A }}{{dt}} = x_A (1 - x_A )(1 - 2 x_A).
\end{equation}
For Eq.(\ref{Eq_xABDO}), there are also three fixed points, as shown in Figure \ref{FigPPGsCGs}(b). $(0, 1)$ and $(1, 0)$ are unstable, and $(0.5, 0.5)$ is stable. These two opinions $A$ and $B$ are equally supported and can coexist in the BDO model.

\begin{figure}[!t]
\centering
\includegraphics[width=5in]{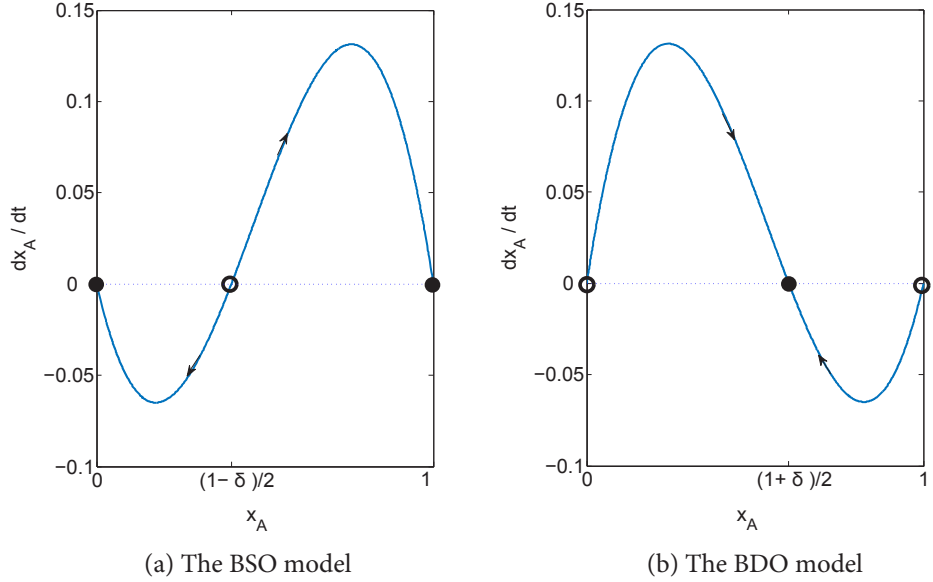}
\caption{Phase diagrams of the binary opinion games in which opinion $A$ is preferred. Black circles represent stable fixed points, and white circles correspond to unstable fixed points. The arrows show the evolutionary direction.}
\label{FigPPGsBOGspreferA}
\end{figure}

In the above given BSO and BDO models, two opinions $A$ and $B$ are treated indiscriminately, the preference for opinion is not taken into consideration. However, in the real world the opinion preference is extensively existent, which is motivated by social prestige, media pressure, and so on. An opinion may be preferred because of accelerating the formation of consensus opinion or inducing the evolutionary direction of opinion dynamics. Formally, the preference for opinion can be reflected on the payoffs. Eq.(\ref{Eq_PT_BSOpreferA}) gives the payoffs in the BSO model with opinion $A$ preferred,
\begin{equation}\label{Eq_PT_BSOpreferA}
\begin{array}{*{20}c}
   {} & {\begin{array}{*{20}c}
   A & \quad B  \\
\end{array}}  \\
   {\begin{array}{*{20}c}
   A  \\
   B  \\
\end{array}} & {\left[ {\begin{array}{*{20}c}
   {1 + \delta } & {0 + \delta }  \\
   0 & 1  \\
\end{array}} \right]}  \\
\end{array}.
\end{equation}
It supposes that an extra profit of $\delta$ will be assigned to opinion $A$ whether it interacts with $A$ or $B$. In this paper, we assume $0 < \delta < 1$. In terms of the RE, the evolutionary formula of $x_A$ is defined by
\begin{equation}\label{Eq_dxApreferA}
\frac{{dx_A }}{{dt}} = x_A (1 - x_A )(2 x_A  - 1 + \delta),
\end{equation}
whose phase diagram is shown in Figure \ref{FigPPGsBOGspreferA}(a). It still contains two stable fixed points of $(0, 1)$ and $(1, 0)$, and an unstable fixed point $(\frac{{1 - \delta }}{2},1 - \frac{{1 - \delta }}{2})$. The opinion preference can not absolutely eliminate the possibility that the population evolves to a non-preferred opinion, but just increases the possibility that the preferred opinion wins. These results are consistent with the social facts.

In the same way, we can analyze the BDO model with opinion preference. The payoff values in the BDO model with opinion $A$ preferred are given in Eq.(\ref{Eq_PT_BDOpreferA}),
\begin{equation}\label{Eq_PT_BDOpreferA}
\begin{array}{*{20}c}
   {} & {\begin{array}{*{20}c}
   A & \quad B  \\
\end{array}}  \\
   {\begin{array}{*{20}c}
   A  \\
   B  \\
\end{array}} & {\left[ {\begin{array}{*{20}c}
   {0 + \delta } & {1 + \delta }  \\
   1 & 0  \\
\end{array}} \right]}  \\
\end{array}.
\end{equation}
where $0 < \delta < 1$. In this case, the evolutionary equation of relative frequency of opinion $A$ is defined by
\begin{equation}\label{Eq_xABDOpreferA}
\frac{{dx_A }}{{dt}} = x_A (1 - x_A )(1 + \delta - 2 x_A).
\end{equation}
According to the corresponding phase diagram as shown in Figure \ref{FigPPGsBOGspreferA}(b), the population eventually evolves to the coexisting state of opinions $A$ and $B$, namely $(\frac{{1 + \delta }}{2},1 - \frac{{1 + \delta }}{2})$. In such stable state of population, the proportion of opinion $A$ is bigger than that of opinion $B$ since opinion $A$ is preferred.

\section{Opinion dynamics with equivocators and preference}\label{SectOGEP}
In the above section, we studied the situation that the opinion of each individual is definitely deterministic, either $A$ or $B$. However, in realistic observations there are lots of equivocators or hedgers whose opinions are indeterminate or ambiguous. The centrist in the vote is a typical example \cite{vazquez2004ultimate,mobilia2011fixation}: in the unknown circumstance individuals will not readily state their positions since they expect to avoid risk. The risk aversion leads to the existence and propagation of equivocators. In the section, apart from the opinion preference, equivocators are also taken into consideration in the evolution of binary opinion.

\subsection{Definition of equivocators}
First, we use a simple model to define the equivocators. Let the distance between two opinions $A$ and $B$ is 1, which implies the similarity of $A$ and $B$ is
\begin{equation}\label{Eq_SAB}
S(A,B) = 1 - D(A,B) = 0,
\end{equation}
where $D(A,B)$ and $S(A,B)$ denote the difference and similarity between both opinions.

In Eq.(\ref{Eq_SAB}), the terms satisfy the symmetry so that $S(A,B) = S(B,A)$. If there is an equivocator, indicated by $E$, who has a distance of $1-r$ away from opinion $A$ and $r$ away from opinion $B$, where $0 < r < 1$, the similarity between $E$ and each binary opinion can be given by
\begin{equation}
S(E,A) = 1 - D(E,A) = r,
\end{equation}
\begin{equation}
S(E,B) = 1 - D(E,B) = 1-r.
\end{equation}
As a result, an equivocator $E$ is defined through these two measures $S(E,A)$ and $S(E,B)$.

\subsection{Case of the BSO model}
\begin{figure}[!t]
\centering
\includegraphics[width=5in]{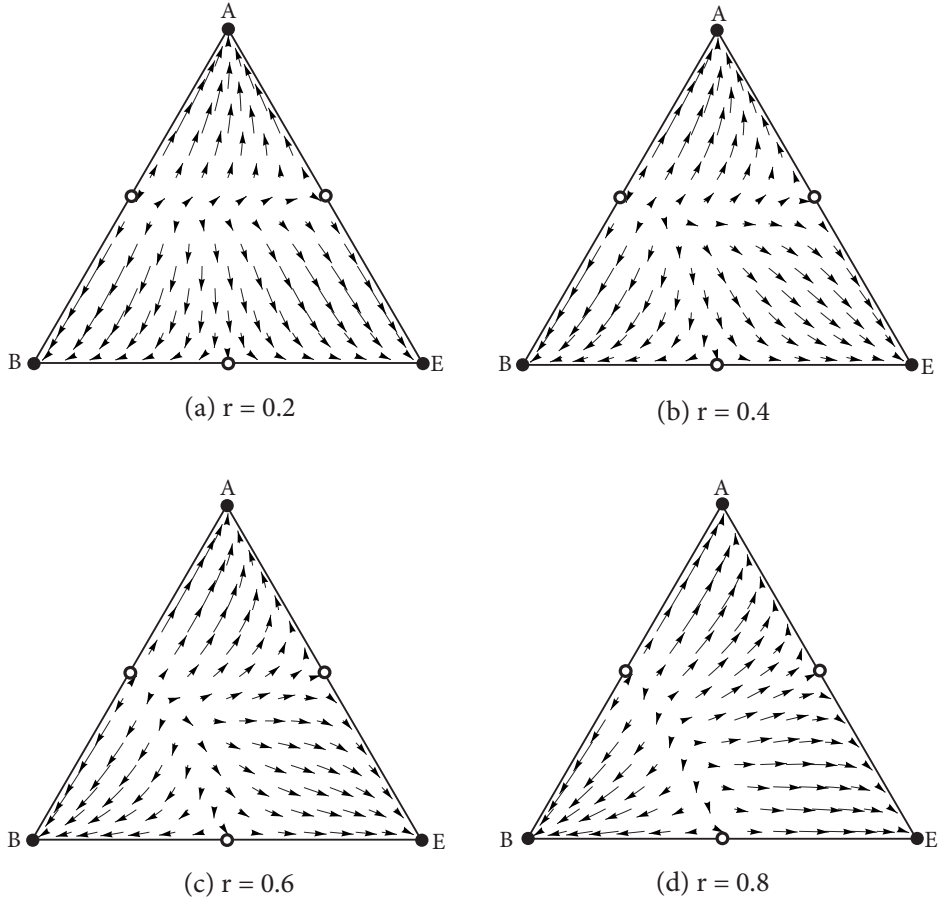}
\caption{Evolutionary dynamics of opinions $A$, $B$, and $E$ in the BSOE model when parameter $r$ takes different values. Black (white) circles are stable (unstable) fixed points.}
\label{EvolDynainBSOE}
\end{figure}

Let us firstly consider the BSO model with equivocators, namely, BSOE model. As above stated, the BSO model rewards the individuals with the same opinion. As an index to measure the similarity between two opinions $P$ and $Q$, the $S(P,Q)$ is naturally appropriate to represent the obtained payoff in the interaction between $P$ and $Q$. Eq.(\ref{Eq_PT_BSOE}) shows the payoffs in the BSOE model, which is a natural extension of Eq.(\ref{Eq_PT_BSO}),
\begin{equation}\label{Eq_PT_BSOE}
\begin{array}{*{20}c}
   {} & {\begin{array}{*{20}c}
   A & \; B \; & \qquad E   \\
\end{array}}  \\
   {\begin{array}{*{20}c}
   A  \\
   B  \\
   E  \\
\end{array}} & {\left[ {\begin{array}{*{20}c}
   1 & 0 & r  \\
   0 & 1 & {1 - r}  \\
   r & {1 - r} & 1  \\
\end{array}} \right]}  \\
\end{array},
\end{equation}
where $0 < r < 1$. Then, the RE is used to investigate the evolution of these opinions. Assume the relative frequency of these opinions is indicated by $x_A$, $x_B$, $x_E$ ($x_A + x_B+ x_E = 1$), respectively.  In terms of the RE, the evolutionary formula is given by
\begin{equation}\label{Eq_dxAxBxE}
\left\{ \begin{array}{l}
 \frac{{dx_A }}{{dt}} = x_A (f_A  - \phi ), \\
 \frac{{dx_B }}{{dt}} = x_B (f_B  - \phi ), \\
 \frac{{dx_E }}{{dt}} = x_E (f_E  - \phi ), \\
 \end{array} \right.
\end{equation}
where $f_A  = x_A  + rx_E$, $f_B  = x_B  + (1-r)x_E$, $f_E  = r x_A  + (1-r) x_B + x_E$, and $\phi  = x_A f_A  + x_B f_B + x_E f_E$. By solving $\frac{{dx_A }}{{dt}} = 0$, $\frac{{dx_B }}{{dt}} = 0$, $\frac{{dx_E }}{{dt}} = 0$, simultaneously, the fixed points $(x_A^*, x_B^*, x_E^*)$ of Eq.(\ref{Eq_dxAxBxE}) are calculated readily. All of fixed points are shown in Table \ref{FPSinBSOE}. According to Theorem \ref{theorem1}, the stability of each fixed point can be found in terms of the associated eigenvalues, which are also shown in Table \ref{FPSinBSOE}. The results show that there are six fixed points, namely $(0,1,0)$, $(0,0,1)$, $(1,0,0)$, $(0.5,0.5,0)$, $(0,0.5,0.5)$, $(0.5,0,0.5)$, which are not concerned with parameter $r$. The first three fixed points are stable, and the last three fixed points are unstable. The evolutionary dynamics of opinions $A$, $B$, and $E$ in the BSOE model can be graphically represented in the simplex, as shown in Figure \ref{EvolDynainBSOE}. Every vertex of the simplex means that there only exists a sole opinion in the population. Edges of the simplex represent that at least one opinion is missing in the population. The interior of the simplex corresponds to the case of all opinions coexistence. At each point of the simplex, the sum of the fractions of these opinions is 100\%. Figure \ref{EvolDynainBSOE} shows that there are three absorbing fixed points. In other words, regardless of the value of parameter $r$ in the BSOE model, the population will eventual evolve to a state which only contains a sole opinion that may be anyone of the opinions $A$, $B$, $E$. The evolutionary process of the population depends on the initial fractions of opinions in the population and parameter $r$. It is impossible for the coexistence of opinions in the BSOE model.

\begin{table}
    \caption{Fixed points and their stability in the BSOE model}\label{FPSinBSOE}
    \begin{center}
    \begin{tabular}{cllllll}
    \hline
    Number   & {Fixed point} &   & Associated eigenvalues &  & Stability &   \\
    \hline
    $p^*_1$   & (0, 1, 0)     &   & $-1$, $-1$, $-r$ &  & stable &   \\
    $p^*_2$   & (0, 0, 1)     &   & $-1$, $-r$, $r-1$ &  & stable &   \\
    $p^*_3$   & (1, 0, 0)     &   & $-1$, $-1$, $r-1$ &  & stable &   \\
    $p^*_4$   & (0.5, 0.5, 0) &   & $-0.5$, $0$, $0.5$ &  & unstable &   \\
    $p^*_5$   & (0, 0.5, 0.5) &   & $\frac{r}{2} - 1$, $\frac{r}{2}$, $r-1$ &  & unstable &   \\
    $p^*_6$   & (0.5, 0, 0.5) &   & $- \frac{r}{2} - \frac{1}{2}$, $\frac{1}{2} - \frac{r}{2}$, $-r$ &  & unstable & \\
    \hline
    \end{tabular}
    \end{center}
\end{table}

\begin{figure}[!t]
\centering
\includegraphics[width=5in]{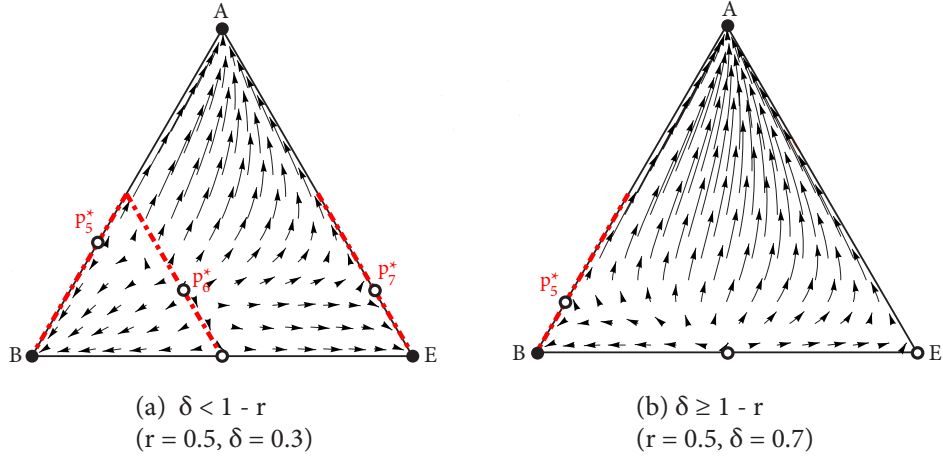}
\caption{Evolutionary dynamics of opinions $A$, $B$, and $E$ in the BSOEP$_A$ model. Black (white) circles are stable (unstable) fixed points. The locations of fixed points $p_5^*$, $p_6^*$, $p_7^*$ change with parameters $r$ and $\delta$, the red dashed lines represent their possible moving trajectory.}
\label{EvolDynainBSOEPA}
\end{figure}

Now, let's turn to  the impact of the opinion preference on the BSOE model. Here we assume that opinion $A$ is preferred in the BSOE model, which is represented by the abbreviation  BSOEP$_A$ model in what follows. The payoff values in such model are given in Eq.(\ref{Eq_PT_BSOEPA}),
\begin{equation}\label{Eq_PT_BSOEPA}
\begin{array}{*{20}c}
   {} & {\begin{array}{*{20}c}
   A & \quad B & \quad E  \\
\end{array}}  \\
   {\begin{array}{*{20}c}
   A  \\
   B  \\
   E  \\
\end{array}} & {\left[ {\begin{array}{*{20}c}
   {1 + \delta } & {0 + \delta } & {r + \delta }  \\
   0 & 1 & {1 - r}  \\
   r & {1 - r} & 1  \\
\end{array}} \right]}  \\
\end{array},
\end{equation}
where $r, \delta \in (0,1)$, which shows an extra profit $\delta$ will be assigned to opinion $A$ no matter which opinion interacts with it. Based on the RE, we can also simulate the evolutionary process of BSOEP$_A$ model in the same way.  All fixed points of the evolutionary dynamics of BSOEP$_A$ model are listed in Table \ref{FPESinBSOEPA}, as well as their associated eigenvalues, existence and stability. Figure \ref{EvolDynainBSOEPA} graphically shows these fixed points, in which red dashed lines mean the associated fixed points are moving with the change of parameters $r$ and $\delta$. Overall, the preference for opinion $A$ increases the opportunity that $A$ becomes the final and only opinion. As shown in Figure \ref{EvolDynainBSOEPA}(a), there are 7 fixed points in the evolutionary dynamics of BSOEP$_A$ model if $\delta < 1 - r$. And the positions of some unstable fixed points change with parameters $r$ and $\delta$. In this situation, $p_1^*$, $p_2^*$, $p_3^*$ are stable, which means any opinion may become the only opinion in the end of evolution. $p_4^*$, $p_5^*$, $p_6^*$, and $p_7^*$ are unstable, which represents that the opinions can not coexist in the BSOEP$_A$ model. Figure \ref{EvolDynainBSOEPA}(b) gives the evolutionary dynamics of these opinions in the BSOEP$_A$ model when $\delta \ge 1 - r$. In that case, there are five fixed points $p_1^*$, $p_2^*$, $p_3^*$, $p_4^*$ and $p_5^*$. Compared with Figure \ref{EvolDynainBSOEPA}(a), $p_1^*$ and $p_2^*$ are still stable, $p_4^*$ and $p_5^*$ are still unstable, while $p_3^*$ changes to unstable case from initially stable point. These results mean that deterministic opinion, either $A$ or $B$, will finally unify the population in the end of evolution, while the opinion dynamics eradicate the equivocators when $\delta \ge 1 - r$  in the BSOEP$_A$ model.

\begin{table}\footnotesize
    \caption{Fixed points and their existence and stability in the BSOEP$_A$ model}\label{FPESinBSOEPA}
    \begin{center}
    \begin{tabular}{cllllll}
    \hline
    Number   & {Fixed point} &   & Associated eigenvalues & Existence  & Stability &   \\
    \hline
    $p^*_1$   & (1, 0, 0)     &   & $-\delta-1$, $-\delta-1$, $r-\delta-1$ & existent & stable &   \\
    $p^*_2$   & (0, 1, 0)     &   & $-1$, $\delta-1$, $-r$ & existent & stable &   \\
    $p^*_3$   & (0, 0, 1)     &   & $-1$, $-r$, $\delta+r-1$ & existent & stable if $\delta < 1 - r$ &   \\
    $p^*_4$   & (0, 0.5, 0.5) &   & $\frac{r}{2} - 1$, $\frac{r}{2}$, $\delta+r-1$ & existent & unstable &   \\
    $p^*_5$   & $(\frac{{1 - \delta}}{2},\frac{{1 + \delta}}{2},0)$ &   & $- \frac{\delta }{2} - \frac{1}{2}$, $\frac{1}{2} - \frac{{\delta ^2 }}{2}$, $- \delta r$ & existent & unstable & \\
    $p^*_6$   & $(\frac{{\delta + r - 1}}{{2r - 2}},\frac{1}{2},\frac{{ - \delta}}{{2r - 2}})$ &   & $\frac{{ - \delta  - 1}}{2}$, $\frac{{r + \delta ^2  - 1 \pm \sqrt {\delta ^4  - 8\delta ^2 r^2  + 10\delta ^2 r - 2\delta ^2  - 8\delta r^3  + 16\delta r^2  - 8\delta r + r^2  - 2r + 1} }}{{4r - 4}}$ & existent if $\delta < 1 - r$ & unstable if existent & \\
    $p^*_7$   & $(\frac{{\delta + r - 1}}{{2r - 2}},0,\frac{{r - \delta - 1}}{{2r - 2}})$ &   & $\frac{{\delta ^2  - r^2  + 2r - 1}}{{2r - 2}}$, $-r$, $\frac{{ - \delta  - r - 1}}{2}$ & existent if $\delta < 1 - r$ & unstable if existent & \\
    \hline
    \end{tabular}
    \end{center}
\end{table}

\subsection{Case of the BDO model}
\begin{figure}[!t]
\centering
\includegraphics[width=5in]{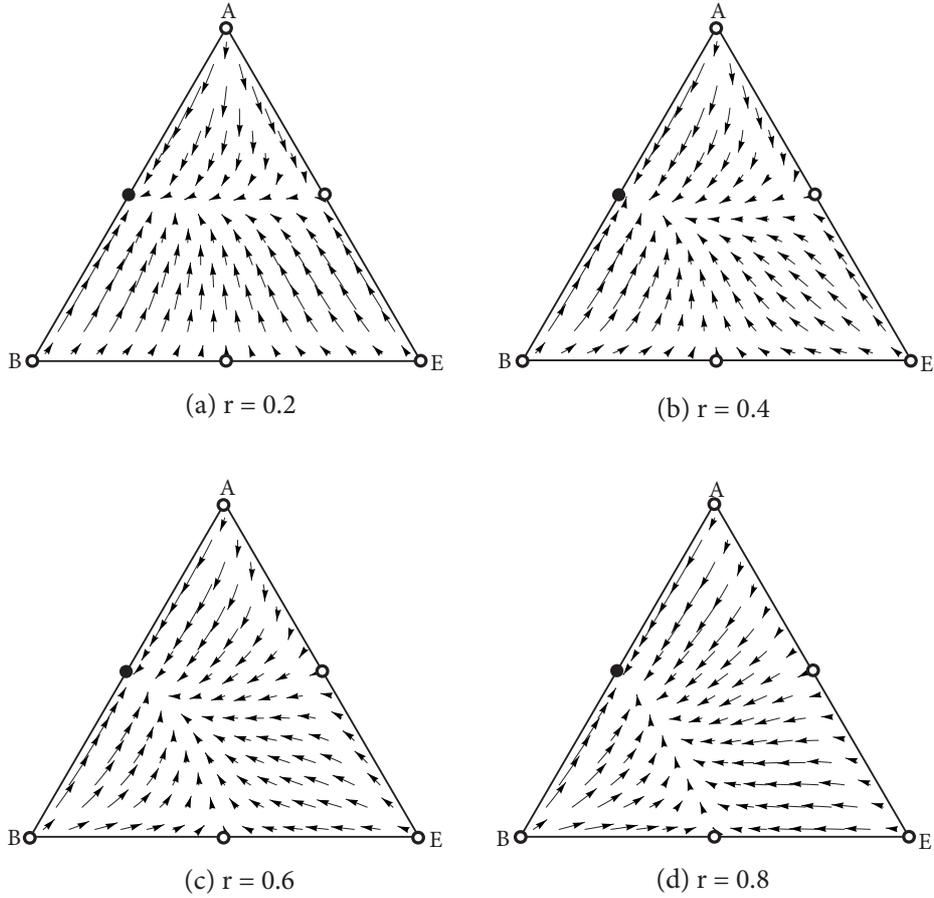}
\caption{Evolutionary dynamics of opinions $A$, $B$, and $E$ in the BDOE model when parameter $r$ takes different values. Black (white) circles are stable (unstable) fixed points.}\label{EvolDynainBDOE}
\end{figure}

In this subsection, we will study the BDO model with the  preference and equivocators. Similar to the above subsection, the BDO model with equivocators, abbreviated as BDOE model, is considered first. Eq.(\ref{Eq_PT_BDOE}) shows the payoffs in the BDOE model,
\begin{equation}\label{Eq_PT_BDOE}
\begin{array}{*{20}c}
   {} & {\begin{array}{*{20}c}
   A & \quad B & \quad  E  \\
\end{array}}  \\
   {\begin{array}{*{20}c}
   A  \\
   B  \\
   E  \\
\end{array}} & {\left[ {\begin{array}{*{20}c}
   0 & 1 & {1 - r}  \\
   1 & 0 & r  \\
   {1 - r} & r & 0  \\
\end{array}} \right]}  \\
\end{array},
\end{equation}
where parameter $r$, $0<r<1$, represents the distance between opinion $E$ and opinion $B$. By means of the EGT and RE, the evolutionary dynamics of BDOE model can be established. In the evolutionary dynamics of BDOE model, there are five unstable fixed points $p_1^*$, $p_2^*$, $p_3^*$, $p_4^*$, and $p_6^*$, and one stable fixed point $p_5^*$, as shown in Table \ref{FPSinBDOE}. The positions of fixed points and their stability are irrelevant to parameter $r$. Figure \ref{EvolDynainBDOE} graphically shows the evolutionary dynamics of opinions $A$, $B$, and $E$ in the BDOE model when parameter $r$ takes different values. These results indicate that in the BDOE model the final state of population is the coexistence of opinions $A$ and $B$, the evolutionary dynamics gradually eliminates the ambiguous opinion $E$ as time increases. In the finally stable state, opinions $A$ and $B$ have the same fraction, which is identical with the classical BDO model. Therefore, the existence of equivocators dose not impact the final result of binary opinions in a circumstance, where players benefit from holding different opinions and the equivocators can not survive.

\begin{table}
    \caption{Fixed points and their stability in the BDOE model}\label{FPSinBDOE}
    \begin{center}
    \begin{tabular}{cllllll}
    \hline
    Number   & {Fixed point} &   & Associated eigenvalues &  & Stability &   \\
    \hline
    1   & (1, 0, 0)     &   & $0$, $1$, $1-r$ &  & unstable &   \\
    2   & (0, 0, 1)     &   & $0$, $r$, $1-r$ &  & unstable &   \\
    3   & (0, 1, 0)     &   & $0$, $1$, $r$ &  & unstable &   \\
    4   & (0.5, 0, 0.5) &   & $r$, $\frac{r}{2} - \frac{1}{2}$ &  & unstable &   \\
    5   & (0.5, 0.5, 0) &   & $-0.5$, $-0.5$, $0$ &  & stable &   \\
    6   & (0, 0.5, 0.5) &   & $1-r$, $- \frac{r}{2}$, $- \frac{r}{2}$ &  & unstable & \\
    \hline
    \end{tabular}
    \end{center}
\end{table}

\begin{figure}[!t]
\centering
\includegraphics[width=5in]{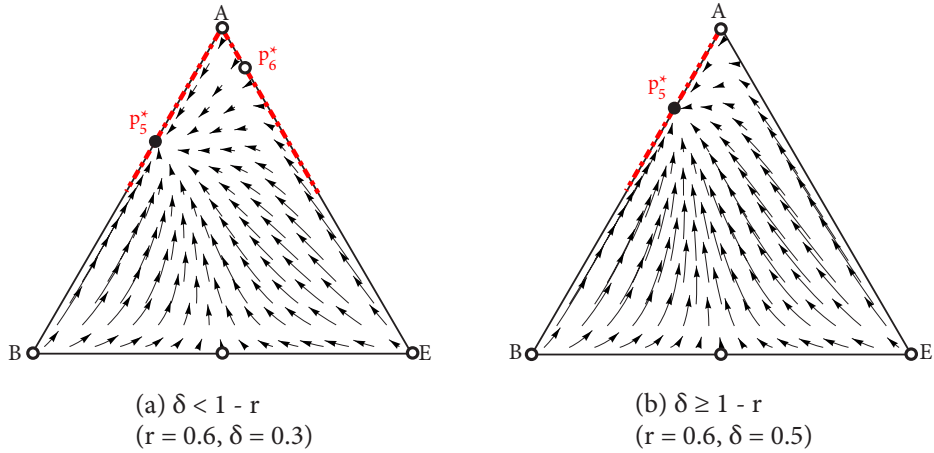}
\caption{Evolutionary dynamics of opinions $A$, $B$, and $E$ in the BDOEP$_A$ model. Black (white) circles are stable (unstable) fixed points. The locations of fixed points $p_5^*$ and $p_6^*$ are changes with parameters $r$ and $\delta$, the red dashed lines represent their possible moving trajectory.}\label{EvolDynainBDOEPA}
\end{figure}

Now, let us add the factor of opinion preference in the BDOE model. The new model is called BDOEP$_A$ model, which means opinion $A$ is preferred in the BDOE model. In the BDOEP$_A$ model, opinion $A$ will obtain an extra profit of $\delta$ when it interacts with any other opinions, as shown in Eq.(\ref{Eq_PT_BDOEPA}),
\begin{equation}\label{Eq_PT_BDOEPA}
\begin{array}{*{20}c}
   {} & {\begin{array}{*{20}c}
   A & \quad B & \quad \qquad E  \\
\end{array}}  \\
   {\begin{array}{*{20}c}
   A  \\
   B  \\
   E  \\
\end{array}} & {\left[ {\begin{array}{*{20}c}
   {0 + \delta } & {1 + \delta } & {1 - r + \delta }  \\
   1 & 0 & r  \\
   {1 - r} & r & 0  \\
\end{array}} \right]}  \\
\end{array},
\end{equation}
where $r, \delta \in (0,1)$. The evolutionary dynamics of BDOEP$_A$ model can be established based on the replicator equation. Table \ref{FPESinBDOEPA} gives all of the fixed points in the evolutionary dynamics of BDOEP$_A$ model, and Figure \ref{EvolDynainBDOEPA} graphically shows these fixed points and their stability. As can be found in Table \ref{FPESinBDOEPA}, there are six fixed points if $\delta < 1 - r$ (see Figure \ref{EvolDynainBDOEPA}(a)); otherwise five fixed points (see Figure \ref{EvolDynainBDOEPA}(b)). In every case, $p_5^*$ is the only stable fixed point, which is an coexistence state of opinions $A$ and $B$. In the final stable state of BDOEP$_A$ model, the fraction of opinions $A$ and $B$ is $\frac{{1 + \delta}}{2}$ and $\frac{{1 - \delta}}{2}$, which is the same with the result of BDO model with opinion $A$ as the preferred one. A preference of $\delta$ for opinion $A$ causes an increase of $\frac{{\delta}}{2}$ to the fraction of opinion $A$ in the population. In addition, according to the results, it can be concluded again that a deterministic opinion is more advantageous than an indeterminate opinion in the circumstance that the diversity of opinions is encouraged.

\begin{table}
    \caption{Fixed points and their existence and stability in the BDOEP$_A$ model}\label{FPESinBDOEPA}
    \begin{center}
    \begin{tabular}{cllllll}
    \hline
    Number   & {Fixed point} &   & Associated eigenvalues & Existence  & Stability &   \\
    \hline
    $p^*_1$   & (0, 1, 0)     &   & $0$, $r$, $\delta+1$ & existent & unstable &   \\
    $p^*_2$   & (0, 0, 1)     &   & $0$, $r$, $\delta-r+1$ & existent & unstable &   \\
    $p^*_3$   & (1, 0, 0)     &   & $1-\delta$, $-\delta$, $1-r-\delta$ & existent & unstable &   \\
    $p^*_4$   & (0, 0.5, 0.5) &   & $- \frac{r}{2}$, $- \frac{r}{2}$, $\delta-r+1$ & existent & unstable &   \\
    $p^*_5$   & $(\frac{{1 + \delta}}{2},\frac{{1 - \delta}}{2},0)$ &   & $- \frac{\delta }{2} - \frac{1}{2}$, $\frac{{\delta ^2 }}{2} - \frac{1}{2}$, $- \delta r$ & existent & stable & \\
    $p^*_6$   & $(\frac{{r - \delta - 1}}{{2r - 2}},0,\frac{{\delta + r - 1}}{{2r - 2}})$ &   & $r$, $\frac{{r^2  - 2r + 1 - {\delta}^2 }}{{2r - 2}}$, $\frac{{ r - \delta - 1}}{2}$ & existent if $\delta < 1 - r$ & unstable if existent & \\
    \hline
    \end{tabular}
    \end{center}
\end{table}

\section{Conclusion}\label{SectConclusion}
This paper has studied the effects of preference and equivocators on binary opinion dynamics based on EGT. Depending on the advantage of capturing the essence of natural selection, the RE is used to act as the role of evolutionary mechanism in the evolution of opinion. Two basic models, BSO and BDO, have been given to describe the interaction of opinions. Based on these two basic models, the impact of opinion preference and equivocators on the evolution of binary opinion is studied respectively and synthetically by using EGT and RE. All equilibrium states and their stability in the binary opinion dynamics have been presented theoretically and mathematically. This work provides a straightforward solution to the binary opinion dynamics.

First, from the simple binary opinion model where preference and equivocators are absent, we show that opinions converges to only one in the BSO model while two opinions can coexist in the BDO model. Then, preference is incorporated and the results show that the factor does not change the existence of equilibrium points and just changes the acceleration and position to the fixed point. Next, equivocators are considered. We find that the opinions converge to only one in the BSOE model and deterministic opinions can coexist (and indeterminate opinion is eliminated) in the BDOE model. Finally, we incorporates preference to the models with equivocators, the opinion dynamics show very interesting results. In this paper, for the simplicity of models, we just consider one level either for preference of opinions or for equivocators, but do not consider the diversity of preference and equivocators. In the future research, the models will be expanded to more realistic and complicated situation.

\section*{Acknowledgment}
The work is partially supported by National High Technology Research and Development Program of China (863 Program) (Grant No. 2013AA013801), R\&D Program of China (2012BAH07B01), National Natural Science Foundation of China (Grant No. 61174022), Specialized Research Fund for the Doctoral Program of Higher Education (Grant No. 20131102130002), the open funding project of State Key Laboratory of Virtual Reality Technology and Systems, Beihang University (Grant No.BUAA-VR-14KF-02), and China Scholarship Council.

\ifCLASSOPTIONcaptionsoff
  \newpage
\fi



%
%
%

\bibliographystyle{IEEEtran}
\bibliography{IEEEabrv,References}

%

%
%
%




\end{document}